\documentclass[preprint,preprintnumbers,showpacs]{revtex4-1}
\usepackage{amsmath}
\usepackage{amsfonts}
\usepackage{amssymb}
\usepackage{bm}
\usepackage{latexsym}
\usepackage{url}
\usepackage{color}

\usepackage{graphicx}
\usepackage{epstopdf}

\begin{document}

\title{Vibrational contribution to the thermodynamics of nanosized precipitates: vacancy-copper clusters in bcc-Fe}

\author{M. Talati}
\email[Email: ]{m.talati@hzdr.de}
\affiliation{Helmholtz-Zentrum Dresden-Rossendorf (HZDR), P.O.Box 510119, 01314 Dresden, Germany}

\author{M. Posselt}
\affiliation{Helmholtz-Zentrum Dresden-Rossendorf (HZDR), P.O.Box 510119, 01314 Dresden, Germany}

\author{G. Bonny}
\affiliation{SCK{$\cdot$}CEN, Boeretang 200, B-2400 Mol, Belgium}

\author{A.T. Al-Motasem}
\affiliation{Helmholtz-Zentrum Dresden-Rossendorf (HZDR), P.O.Box 510119, 01314 Dresden, Germany}

\author{F. Bergner}
\affiliation{Helmholtz-Zentrum Dresden-Rossendorf (HZDR), P.O.Box 510119, 01314 Dresden, Germany}

\date{\today}


\begin{abstract}
Within the harmonic approximation, the effects of lattice vibration on the thermodynamics of nano-sized coherent clusters in bcc-Fe consisting of vacancies and/or copper are investigated. A combination of on-lattice simulated annealing based on Metropolis Monte Carlo simulations and off-lattice relaxation by Molecular Dynamics is applied to obtain the most stable cluster configurations at T = 0 K. The most recent interatomic potential built within the framework of the embedded atom method for the Fe-Cu system is used. The vibrational part of the total free energy of defect clusters in bcc-Fe is calculated using their phonon density of states. The total free energy of pure bcc-Fe and fcc-Cu as well as the total formation free energy and the total binding free energy of the vacancy-copper clusters are determined for finite temperatures. Our results are compared with the available data from previous investigations performed using empirical many-body interatomic potentials and first-principle methods. For further applications in rate theory and object kinetic Monte Carlo simulations, the vibrational effects evaluated in the present study are included in the previously derived analytical fits based on the classical capillary model.
\end{abstract}



\maketitle

\section{Introduction}

The level of impurity copper in reactor pressure vessel (RPV) steels of existing fission reactors (0.03 to 0.3 wt\%) is typically higher than the solid solubility of Cu in Fe at RPV operation temperature~\cite{miller1998,auger2000,perez2005}. Neutron irradiation gives rise to a steady-state vacancy concentration that is orders of magnitude higher than in thermal equilibrium and enables Cu to diffuse efficiently via a vacancy mechanism~\cite{odette1998,christien2004}. As a consequence, Cu-rich clusters form and take up a number of vacancies. In fact, mixed Cu-vacancy clusters with non-monotonically varying fractions of Cu and vacancies were observed recently in binary Fe-Cu alloys~\cite{xu2006,nagai2003,bergner2010}. Similar processes take place in RPV steels, although other impurity and alloying elements may also play a role and vacancies seem to contribute less. Cu-rich clusters or precipitates in both binary Fe-Cu alloys (Cu \textless 0.3 wt\%) and RPV steels are typically smaller than 4 nm in diameter~\cite{ulbricht2007,ulbricht2006,bergner2010}. Experimental investigations~\cite{othen1994,pizzini1990} also show that these Cu-rich precipitates are coherent i.e. they possess the bcc-structure of iron.

Irradiation-enhanced formation and evolution of the Cu-rich nano clusters or precipitates are multiscale processes. These processes involve various physical phenomena and can be efficiently investigated with rate theory ~\cite{christien2004,mathon1997} and object kinetic Monte Carlo (OKMC) simulations~\cite{domain2004,caturla2006} covering more or less realistic time and length scales. Important thermodynamic parameters, in particular the binding free energy of monomers (Cu-atoms or vacancies) to defect clusters required for these coarse-grained or continuum approaches are, however, hardly accessible from experiments. But atomistic computer simulations can provide these parameters.

In this study, our main objective is to quantitatively determine the vibrational contribution to the thermodynamics of defect clusters consisting of vacancies and/or Cu-atoms in the Fe-Cu system with cluster size larger than that first principle methods can treat. For this purpose, the most recent interatomic potential for the Fe-Cu alloy is employed~\cite{pasianot2007}. We perform here a combination of Metropolis Monte Carlo (MMC) simulations and Molecular Dynamics (MD) simulations to find the energetically most stable defect configurations. This method is previously employed by Al-Motasem et al.~\cite{almotasem2011} and is similar to the procedures used by Takahashi et al.~\cite{takahashi2003} and Kulikov et al.~\cite{kulikov2006}. For small Cu-clusters (up to 15 Cu-atoms) in bcc-Fe, first principle studies using density functional theory (DFT) reveal that the vibrational effect is comparable ($\sim$36$\%$) to configurational entropy~\cite{yuge2005}. To determine the vibrational effects, we find the phonon frequencies of the stable structures under the harmonic approximation, with the computation of thermodynamic quantities following. The total formation free energy and the total binding free energy for pure and mixed defect clusters are calculated. Moreover, in order to furnish the input parameters for further applications in rate theory and OKMC simulation, we also aim here to determine simple but practically correct analytical fit formulae that render the total and the monomer binding free energy of defect clusters for any finite temperatures. The computational procedure is detailed in section 2. Results are presented and discussed in section 3, followed by the summary and conclusion of this work.

\section{Computational Method} 

In order to find the most stable configurations of coherent clusters consisting of $\textit l$ vacancies and $\textit m$ Cu atoms (i.e. $v_l Cu_m $ clusters), at first we performed on-lattice simulated annealing based on Metropolis Monte Carlo (MMC) simulation as described in ref.~\cite{almotasem2011}. The initial configurations contained randomly-distributed isolated vacancies and Cu atoms at 600 K. Then, we decreased the temperature to 0 K reducing it by 60 K at every 100 MMC steps. A single cluster is found at 0 K for all cases considered in this study. Subsequently, we performed a quasi-dynamic quench through Molecular Dynamics simulation at 0 K to relax the positions of Fe and Cu atoms off-lattice following the procedure reported in ref.~\cite{almotasem2011}. The lowest formation energy, $\textit E_f(v_\textit l Cu_\textit m)$ is then obtained by performing several such relaxations for slightly different values of the lattice constant. This quantity is defined as 
\begin{equation}
E_f(v_l Cu_m) = E(v_l Cu_m)-(N-l-m)E^{coh}_{Fe}-mE^{coh}_{Cu}
\label{eq:1}
\end{equation} 
where $N$ is the total number of regular lattice sites. $E(v_l Cu_m)$ is the configurational energy of the simulation box containing the most stable configurations of the $v_l Cu_m$ cluster. $E^{coh}_{Fe}$ and $E^{coh}_{Cu}$ represent the cohesive energy per atom of bcc-Fe and fcc-Cu, respectively ($E^{coh}_{Fe}$ = -4.122 eV and $E^{coh}_{Cu}$ = -3.540 eV for chosen interatomic potential) which are chosen as reference systems. The total binding energy of a $v_lCu_m$ cluster, $E_{bind}(v_l Cu_m)$ describes the energy change when isolated vacancies and Cu atoms are combined to form a cluster and is defined by
\begin{equation}
E_{bind}(v_l Cu_m) = E_f(v_l Cu_m)-lE_f(v_1)-mE_f(Cu_1),
\label{eq:2}
\end{equation}
where $E_f(v_1)$ and $E_f(Cu_1)$ are the formation energy of a monovacancy and of a single Cu atoms, respectively (cf. Eq. (\ref{eq:1})). $E_f(Cu_1)$ is also called substitutional energy. Negative values of $E_{bind}(v_l Cu_m)$ suggest an energetically favorable condition for the monomers to form a cluster. We employed here, the most recent and currently the most suitable empirical many-body interatomic potential of Pasianot and Malerba (PM)~\cite{pasianot2007} developed for the description of radiation effects in Fe-Cu binary alloys. It is based on the Mendelev~\cite{mendelev2003} and the Mishin~\cite{mishin2001} potentials for pure Fe and Cu, respectively. The simulation box of bcc-Fe lattice is chosen to be a cubic in shape with each edge of 10${a}$ length, where ${a}$ is the lattice constant. Three dimensional periodic boundary conditions are applied. In this study, pure clusters containing up to 80 either vacancies or copper atoms are considered. Maximum cluster size amounts to total of 100 vacancies and copper atoms for mixed clusters. In agreement with experimental findings~\cite{othen1994,pizzini1990}, all the clusters considered in this paper are assumed to possess the bcc-structure of iron. 

The phonon frequencies $\omega_i$ are determined by diagonalizing the dynamical matrix under the harmonic approximation for the most stable configurations obtained at 0 K. At $\Gamma$-point, the dynamical matrix $\Phi_{ij}$ is defined as 
\begin{equation}
\Phi_{ij} = -(\frac{1}{\sqrt{m_im_j}}\frac{\partial \bf F_i}{\partial \bf r_j}),
\label{eq:3}
\end{equation} where $m_i$ is the mass of $i^{th}$ atom. It is calculated by applying small displacements $r_j$ away from their most stable atomic positions and evaluating the subsequent induced forces (${\bf F_i}$) and force derivatives (${\frac{\partial \bf F_i}{\partial \bf r_j}}$). 

The phonon calculations allow us to compute various thermodynamic properties, e.g. free energy, entropy, heat capacity etc. The vibrational contribution to the total free energy (from now on the vibrational free energy) of a solid, $G^{vib}$ at finite temperature T is calculated using the following expression: 
\begin{equation}
G^{vib}(V,T) = U^{vib}(V,T)-TS^{vib}(V,T),
\label{eq:4}
\end{equation}
where $U^{vib}(V,T)$ represents the vibrational internal energy and $S^{vib}(V,T)$ the vibrational entropy. In the harmonic approximation, these contributions (cf. Eqs.~\ref{eq:5},~\ref{eq:6})) are represented as the sum of individual normal frequencies, $\omega_i$. The total number of individual normal frequencies are three times the total number of atoms in the primitive cells. Because of the translational invariance of the system three eigen values corresponding to the translational degrees of freedom are zero and hence,

\begin{equation}
U^{vib}(V,T) = \sum^{3N-3}_{i=1}\left[\frac{\hbar\omega_i}{e^{\frac{\hbar\omega_i}{k_BT}}-1} + \frac{1}{2}{\hbar\omega_i}\right]
\label{eq:5}
\end{equation} and  

\begin{equation}
S^{vib}(V,T) = k_B\sum^{3N-3}_{i=1}\left[\frac{\hbar\omega_i}{k_BT}\left({e^{\frac{\hbar\omega_i}{k_BT}}-1}\right)^{-1} - ln\left(1-{e^{-\frac{\hbar\omega_i}{k_BT}}}\right)\right].
\label{eq:6}
\end{equation} Here, $k_B$ represents the Boltzmann constant.

An alternative way to calculate the vibrational free energy ($G^{vib}$) here, is through the phonon density of states (PDOS), $g(\omega)$:

\begin{equation}
G^{vib}(V,T) = (3N-3)\int^{\infty}_{0}k_BT\:ln\left(2sinh \left(\frac{\hbar\omega}{2k_BT}\right)\right)g(\omega)d\omega,
\label{eq:7}
\end{equation}
where $g(\omega)$ is the normalized so that 

\begin{equation}
\int^{\infty}_{0}
	 g(\omega)d\omega = 1.
\label{eq:8}
\end{equation} The phonon density of states $g(\omega)$, which can also be considered as a histogram of phonon frequencies with vanishing bin size, gives the number of vibrational modes per phonon frequency. 

The vibrational part of total formation free energy of a $v_l Cu_m$ cluster is calculated from the following expression:
\begin{equation}
G^{vib}_f(v_l Cu_m,V,T) = G^{vib}(v_l Cu_m,V,T) - (N-l-m)G^{vib}_{coh}(Fe,V,T) - mG^{vib}_{coh}(Cu,V,T),
\label{eq:9}
\end{equation} 
where, $G^{vib}(v_l Cu_m,V,T)$ determined using Eq.~(\ref{eq:7}) represents the vibrational free energy of the system with the $v_l Cu_m$ cluster. $G^{vib}_{coh}(Fe,V,T)$ and $G^{vib}_{coh}(Cu,V,T)$ represent the vibrational contribution to the cohesive energy per atom at given temperature (T) and volume (V) of bcc-Fe and fcc-Cu, respectively. The total formation free enery is then written as  \begin{equation}
G^{Total}_f(v_l Cu_m,V,T)= E_f(v_l Cu_m) - TS^{conf}_f(v_lCu_m) + G^{vib}_f(v_l Cu_m,V,T),
\label{eq:10}
\end{equation} 
where $S^{conf}_f$ (= $k_B lnW$) is the formation entropy at T = 0 K which is determined by the number W of different three-dimensional arrangements of a cluster with a given shape at a given position. In the present paper the contribution of $S^{conf}_f(v_lCu_m)$ is neglected since the computation time is prohibitively long. The efficient determination of $S^{conf}_f(v_lCu_m)$ is a subject of ongoing investigation.   

The vibrational part of total binding free energy and the total binding free energy can be obtained in a similar fashion from the following formulae,
\begin{equation}
G^{vib}_{bind}(v_l Cu_m,V,T) = G^{vib}_f(v_l Cu_m,V,T) - lG^{vib}_f(v_1,V,T) - mG^{vib}_f(Cu_1,V,T)
\label{eq:11}
\end{equation}and
\begin{equation}
G^{Total}_{bind}(v_l Cu_m,V,T) = E_{bind}(v_l Cu_m) + G^{vib}_{bind}(v_l Cu_m,V,T)
\label{eq:12}
\end{equation}
where $G^{vib}_f(v_l Cu_m,V,T)$ , $G^{vib}_f(v_1,V,T)$ and $G^{vib}_f(Cu_1,V,T)$ represent the vibrational part of free formation energy of the $v_l Cu_m$ cluster, of a monovacancy and a single Cu in bcc-Fe, respectively (cf. Eq.~(\ref{eq:10})). $E_{bind}(v_l Cu_m)$ represents the binding energy of a cluster at 0 K (cf. Eq.~(\ref{eq:2})). 

The monomer binding free energy is a binding energy of a monovacancy or of a single Cu atom to a pure $v_l$ or $Cu_m$ cluster. This quantity, at T $\neq$ 0 and zero pressure is required as input parameters for rate theory and object kinetic Monte Carlo simulations and can be determined from the following expression:

\begin{eqnarray}
&& G^{Total}_b(v,v_l,V,T) = G^{Total}_{bind}(v_l,V,T) - G^{Total}_{bind}(v_{l-1},V,T)\nonumber\\
&& G^{Total}_b(Cu,Cu_m,V,T) = G^{Total}_{bind}(Cu_m,V,T) - G^{Total}_{bind}(Cu_{m-1},V,T)
\label{eq:13}
\end{eqnarray}

\section{Results and Discussion}

\subsection{Phonon density of states and vibrational free energy of bcc-Fe and fcc-Cu}

Figure 1 depicts the phonon density of states (PDOS) of pure bcc-Fe and fcc-Cu, which are in good agreement with the available experimental observation and theoretical calculations~\cite{fultz2010,bogdanoff2003,bogdanoff1999,reith2009}. The PM potential which is based on the Mendelev potential~\cite{mendelev2003} for pure Fe and on the Mishin potential~\cite{mishin2001} for pure Cu seems to reproduce the experimental phonon density of states of elemental bcc-Fe and fcc-Cu well.
\begin{figure}[ht]
  \centering
	\includegraphics[width=8cm]{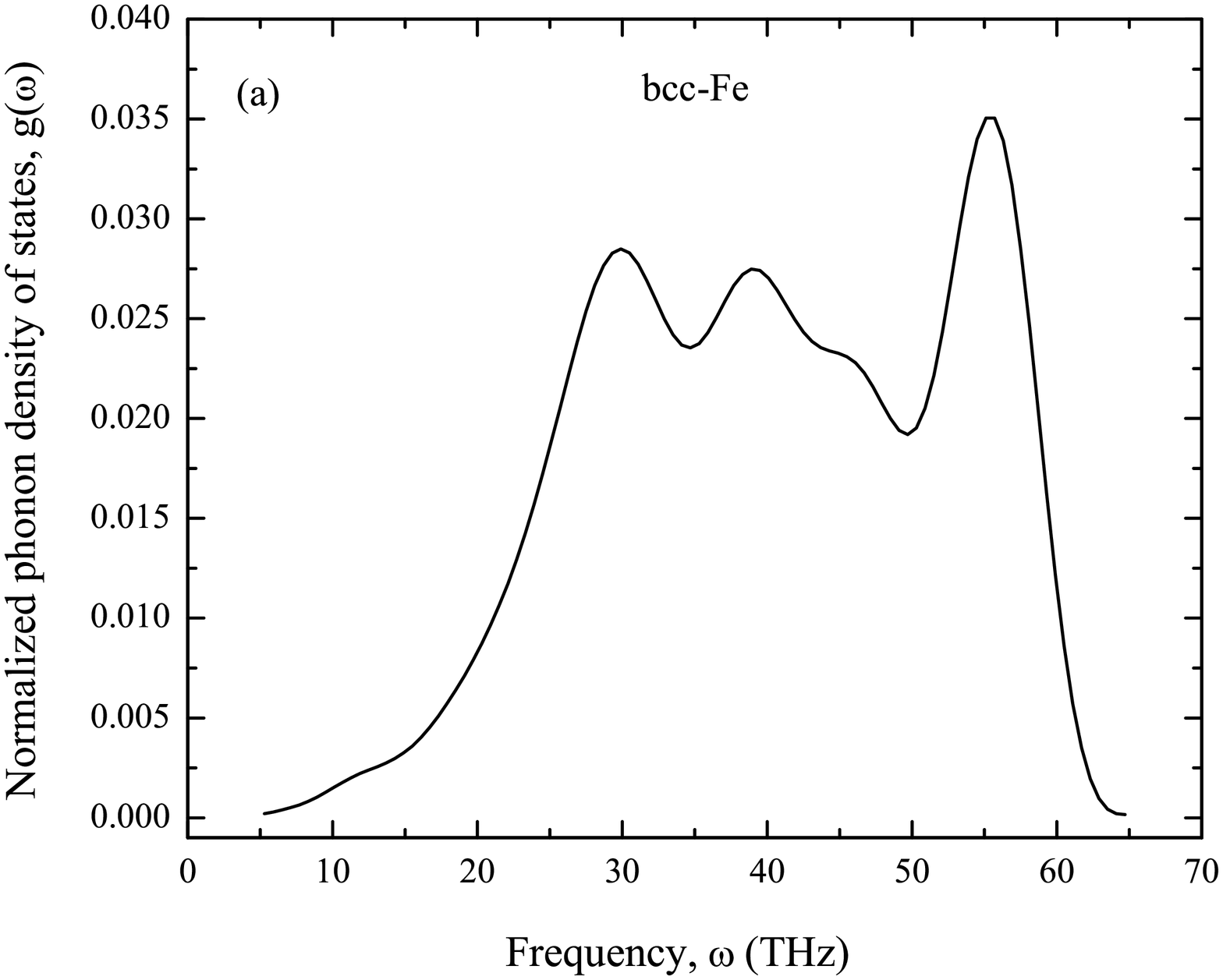}\quad
  \includegraphics[width=8cm]{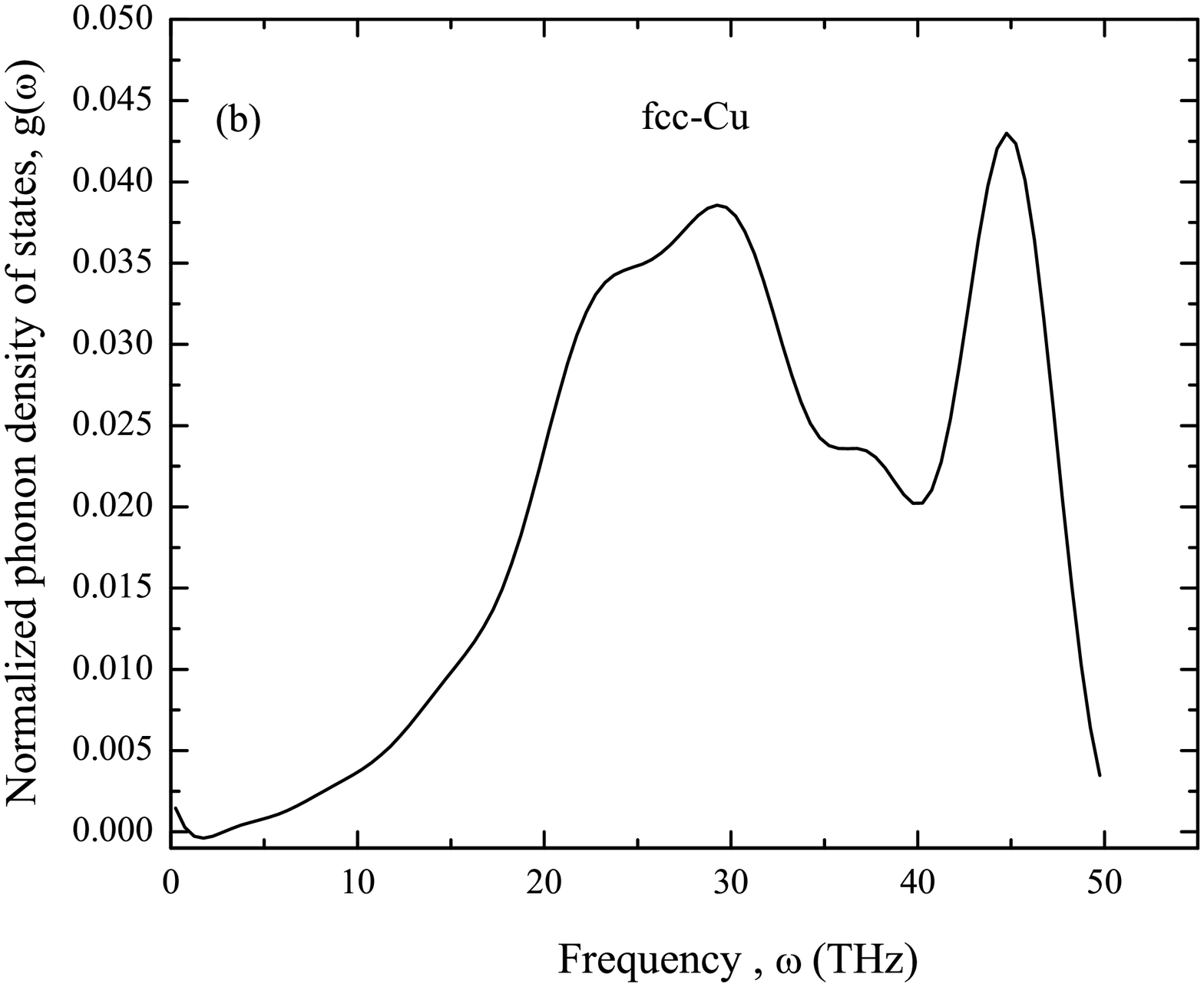}
	\caption{\label{figs:Fig.1}}Phonon density of states of (a) bcc-Fe and (b) fcc-Cu.
\end{figure}
The vibrational free energy of the considered systems is then calculated using their PDOS in Eq.~(\ref{eq:7}). As shown in Fig. 2, our results for bcc-Fe and fcc-Cu are compared with the Scientific Group Thermodata Europe (SGTE) database and results from other empirical potentials for these two elements~\cite{dinsdale1991,koermann2008,engin2008,ludwig1998,caro2006}. The SGTE methodology includes CALPHAD calculations and experimental data and provides the thermodynamic data for inorganic and metallurgical systems. The CALPHAD data in ref.~\cite{koermann2008} has been obtained with the THERMOCALC program and the SGTE unary database~\cite{guillemet1985}. The vibrational free energy of bcc-Fe and fcc-Cu calculated in the present work fall very much into the range of available literature data. At 1200 K, the calculated vibrational free energy of bcc-Fe using the PM potential is about 160 meV and 100 meV higher than SGTE database~\cite{dinsdale1991} and CALPHAD data~\cite{koermann2008}, however, it falls within 50 meV range of vibrational free energy calculated using other empirical potentials such as the Meyer-Entel potential, the Johnson potential which is a recent update of the Johnson-Oh potential, and the Finnis-Sinclair potential~\cite{engin2008}. At around 600K - an operating temperature of a nuclear reactor, the former difference in the vibrational free energy is about 70 meV and 30 meV, respectively while the latter is within 20 meV range. The Meyer-Entel potential predicts the highest values of the vibrational free energy and hence the farthest from SGTE database or CALPHAD data while the PM potential provides the closest values. Similarly, though the Ludwig-Farkas potential predicts the vibrational free energy of the fcc-Cu (in~\cite{caro2006}) closest to the SGTE database~\cite{dinsdale1991}, the PM potential calculates it within the closer range of 30 meV and 10 meV at 1200 K and 600 K respectively. The ability to reproduce the experimental PDOS of bcc-Fe and fcc-Cu to a satisfactorily extent and to predict their vibrational free energy close to the SGTE database, the choice of the PM potential is justified for the considered systems in the present work.
\begin{figure}[ht]
  \centering
	\includegraphics[width=8cm]{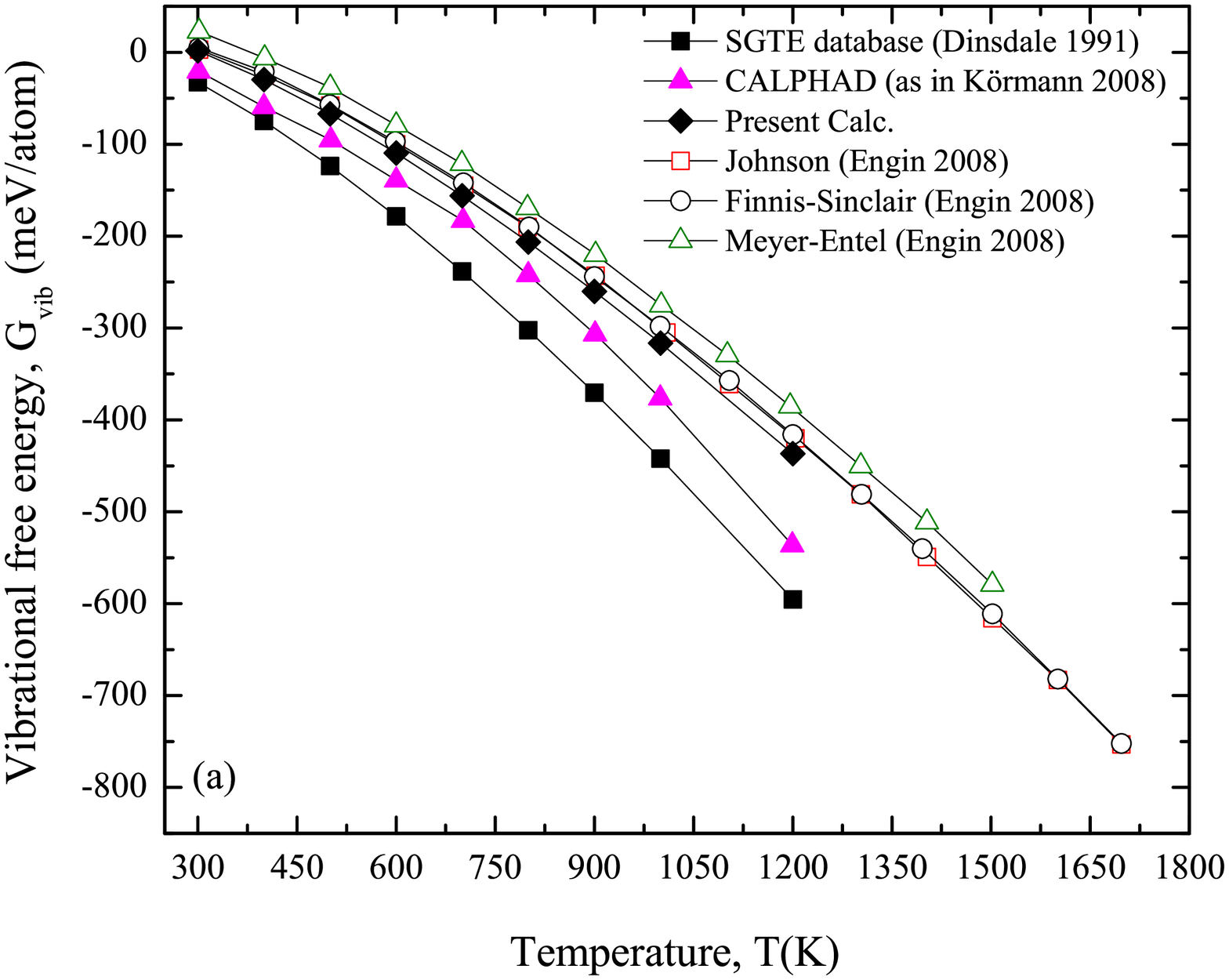}\quad
	\includegraphics[width=8cm]{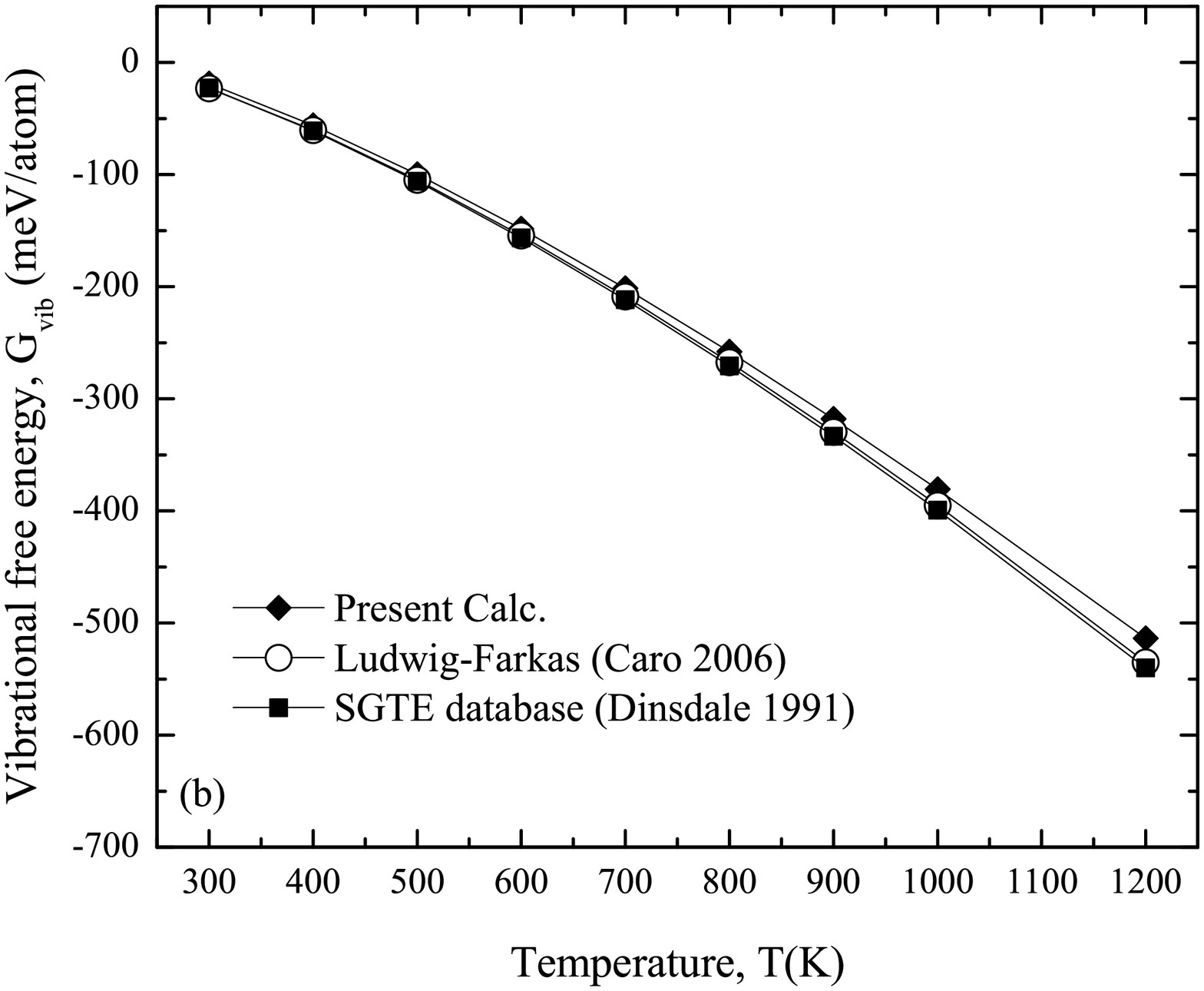}
	\caption{\label{figs:Fig.2}Vibrational free energy (meV/atom) of (a) bcc-Fe and (b) fcc-Cu. Present results with the PM potential are compared with the SGTE or CALPHAD data and the other potentials for bcc-Fe \cite{dinsdale1991,koermann2008,engin2008} and for fcc-Cu \cite{dinsdale1991,caro2006}.}
\end{figure}

\subsection{Formation free energy of the clusters}
The presence of considered number of vacancies and substitutional Cu-atoms seems to very weakly affect peak-positions and form of the PDOS of bcc-Fe to notice any visible difference. For this reason, the PDOS of defect clusters in bcc-Fe are not shown here. First principle study for a monovacancy in bcc-Fe also reports a similar observation with a slight phonon spectrum-shift towards lower frequencies~\cite{lucas2009}. Figure 3 depicts the total formation free energy of a monovacancy and a single Cu atom in bcc-Fe. In the present study, the values of formation energy of a monovacancy and of a single Cu atom are $E_f(v_1)$ = 1.710 eV and $E_f(Cu_1)$ = 0.4369 eV, respectively which are obtained using the PM potential~\cite{almotasem2011}. Due to an increase in vibrational interactions (or thermal excitations) with temperature, the probability to form vacancies in a lattice also increases. In other words, the vibrational contributions lower the total formation free energy of a monovacancy ($G^{Total}_f(v_1,T)$) and of pure vacancy clusters ($G^{Total}_f(v_l,T)$) at higher temperatures as shown in Fig. 3(a) and Fig. 5(a), respectively. Recent computational investigations on the monovacancy in bcc-Fe with both first principle calculations~\cite{lucas2009} and many-body empiricial potential~\cite{mendelev2009} also conform to our results for a monovacancy case, and hence for pure vacancy clusters ($v_l$) in general. However, it is worth to note that the formation energy of a monovacancy in bcc-Fe estimated from experimental investigations~\cite{ehrhart1991} and using other empirical potentials~\cite{mendelev2009,ackland1997,kislov2003,pohlong1998} are slightly different from the first principle calculations~\cite{lucas2009,sato2009,domain2001,domain2006,ohnuma2009,soisson2007,tateyama2003}. Mendelev et al.~\cite{mendelev2009} estimated and interpolated the temperature dependence of total formation free energy for a monovacancy in bcc-Fe. They fitted interpolation coefficients to MD data at low temperatures and the equilibrium defect concentrations at high temperatures. These coefficients seem to capture additional anharmonic effects leading to a further decrease of $G^{Total}_f(v_1 ,T)$ above 900 K as compared to our results. Nevertheless, in agreement with the available data our results for the monovacancy case provided us confidence to further the similar computation for pure vacancy clusters ($v_l$) of larger size. The decrease in total formation free energy of pure $v_l$ clusters, demonstrated in Fig. 5(a) as a function of temperature, indicates an enhanced activity of vacancy cluster formation or vacancy trapping in bcc-Fe at higher temperatures.

\begin{figure}[ht]
  \centering
	\includegraphics[width=8cm]{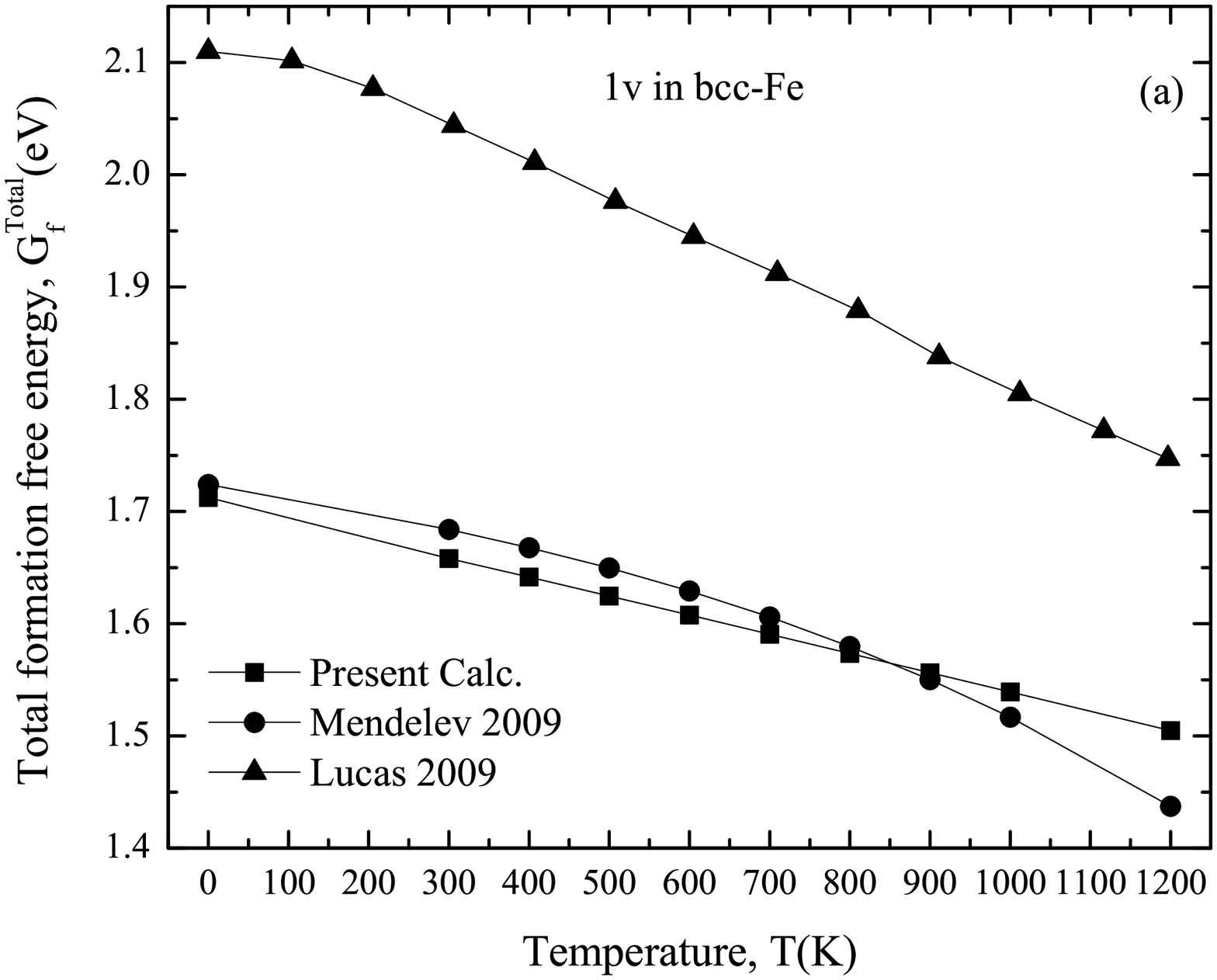}\quad
	\includegraphics[width=8cm]{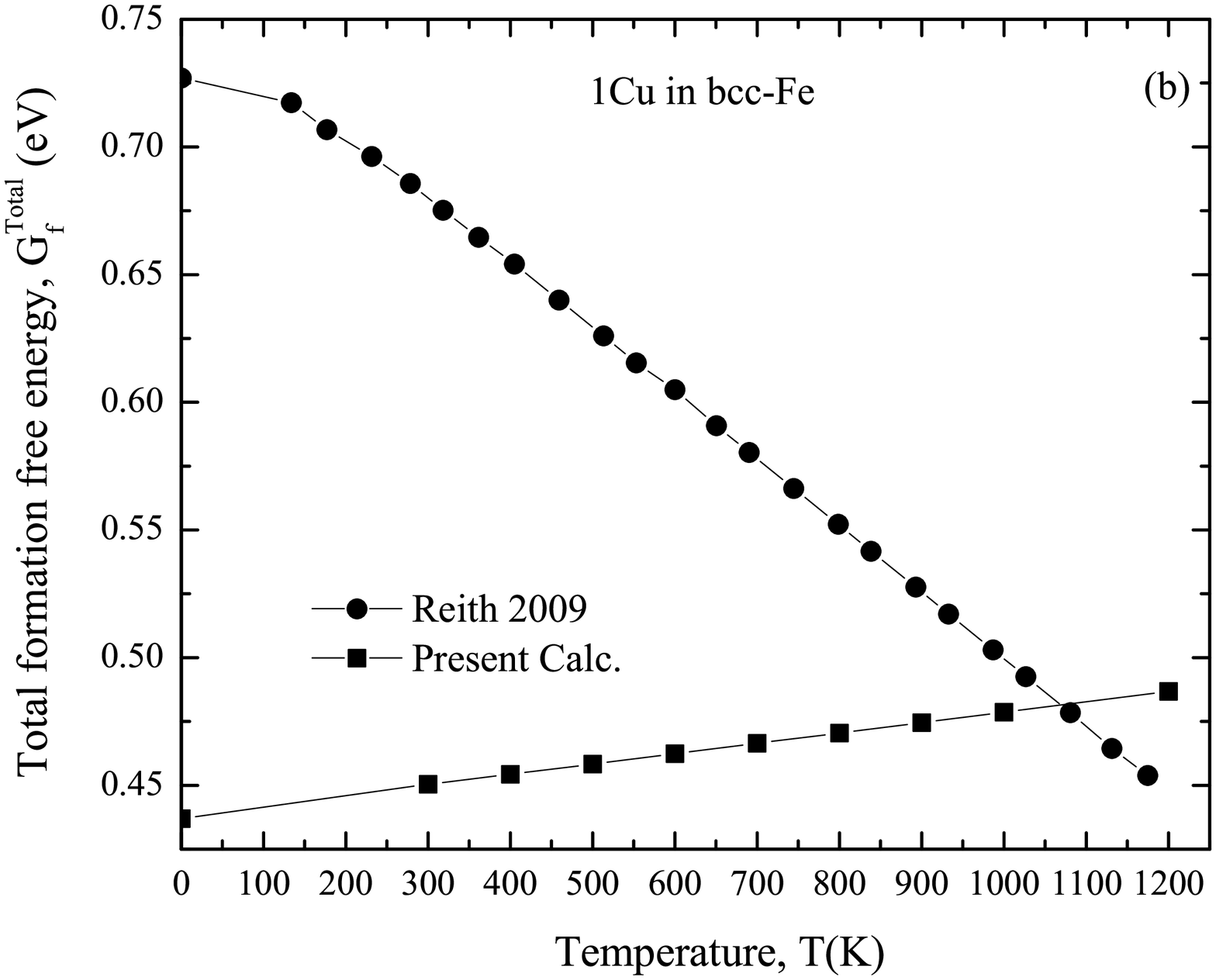}
	\caption{\label{figs:Fig.3}Total formation free energy (eV) of (a) a monovacancy in bcc-Fe and (b) a single Cu atom in bcc-Fe.}
\end{figure}

Our results for total formation free energy of a single Cu-atom in bcc-Fe, on the other hand is contrary to our expectation and the prediction of first principle calculations~\cite{reith2009}(see Fig. 3(b)). First principle calculation using the VASP code predicts the formation energy of a single Cu in bcc-Fe, $E_f(Cu_1)$ = 0.55-0.77 eV~\cite{vincent2005,domain2001,domain2006,reith2009} while the SIESTA code predicts $E_f(Cu_1)$ = 0.48 eV~\cite{soisson2007}. The rigid lattice model~\cite{soisson2007} estimates $E_f(Cu_1)$ = 0.545 eV. The PM potential fitted against the experimental values of Cu solubility in bcc-Fe~\cite{perez2005,salje1978} and neglecting the vibrational contribution estimates the lower formation energy, $E_f(Cu_1)$ = 0.4369 eV as compared to the reported DFT values~\cite{domain2001,domain2006,reith2009,vincent2005}. In Fig. 3(b) we compare the temperature dependence of the total formation free energy of a single Cu-atom in bcc-Fe with the VASP calculations of Reith et al.~\cite{reith2009}. The values of vibrational part of total formation free energy ($G^{vib}_f(Cu_1,V,T)$) as reported by Reith et al. are negative and the absolute values of $G^{vib}_f(Cu_1,V,T)$ increase with temperatures. As shown in Fig. 3(b) the total formation free energy, $G^{Total}_f(Cu_1,V,T)$ remains positive and decreases with temperature where the relatively high formation energy ($E_f(Cu_1)$ = 0.77 eV) is summed with negative values of vibrational part of total formation free energy ($G^{vib}_f(Cu_1,V,T)$) according to Eq.~(\ref{eq:10}). The PM potential, on the other hand renders the positive values of $G^{vib}_f(Cu_1,T)$ with its absolute values increasing with temperature. The calculated total formation free energy as demonstrated in Fig. 3(b) increases with temperature unlike the case reported by Reith et al. Although the increase in total formation free energy is small ($\sim$ 0.05 eV at 1200 K), it indicates a slight unfavourable condition for the formation a single Cu-atom in bcc-Fe at higher temperatures. A full thermodynamic integration~\cite{koning1996,koning2005} being consistent with the harmonic approximation in the present study, also qualitatively predicts a similar trend of total formation free energy of a single Cu in bcc-Fe as a function of temperature. 

The solubility limit of Cu in iron or Cu concentration (in at \%) in iron at equilibrium, $C^{Sol}_{Cu}$ is related to the total formation free energy by
\begin{equation}
G^{Total}_f(Cu_1,V,T) = -k_BTln(C^{Sol}_{Cu}). 
\label{eq:14}
\end{equation}
Figure 4 shows experimental and theoretical solubility data from the literature and they are compared with the results obtained from Eq.~(\ref{eq:14}). Experimental data were obtained from measuring the concentration and diffusion profile of a thin Cu layer deposited on an iron crystal~\cite{salje1977}; thermoelectric power and small angle X-ray scattering measurements in thermally aged pure Fe-Cu alloys~\cite{perez2005}. The computational data were obtained using an advanced Monte Carlo method implemented in the Alloy Theoretic Automated Toolkit package (ATAT)~\cite{vandewalle2002} with the PM potential denoted as CO5.20~\cite{pasianot2007}. Note that in both experimental and theoretical investigations the solubility was determined for the thermodynamic phases, fcc-Cu and bcc-Fe in equilibrium. Our results for solubility limits of Cu in bcc-Fe with the PM potential neglecting the vibrational effects i.e. corresponding to the calculations at T = 0 K formation energy are in a good agreement with experimental and computational data~\cite{salje1977,perez2005,pasianot2007}. Inclusion of vibrational effects at T $\neq$ 0 K, however, lowers the Cu solubility limit. It is not clear to us whether the reduction in solubility limits of Cu in bcc-Fe calculated using the PM potential including the vibrational effects is real or an artefact of the employed potential.
\begin{figure}[ht]
  \centering
	\includegraphics[width=8cm]{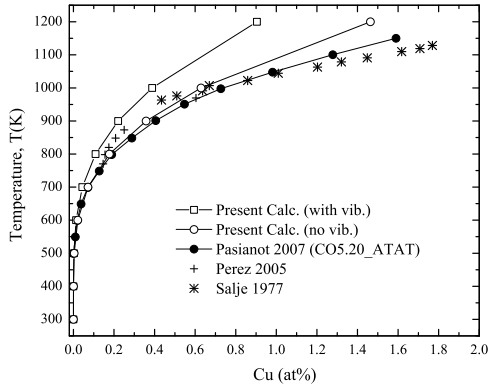}\quad
	\caption{\label{figs:Fig.4}Solubility limit of Cu in bcc-Fe. Present results are compared with experiments~\cite{salje1977,perez2005} and calculations performed with the ATAT package~\cite{vandewalle2002} using the PM potential denoted as CO5.20~\cite{pasianot2007}.}
\end{figure}

\begin{figure}[ht]
  \centering
	\includegraphics[width=8cm]{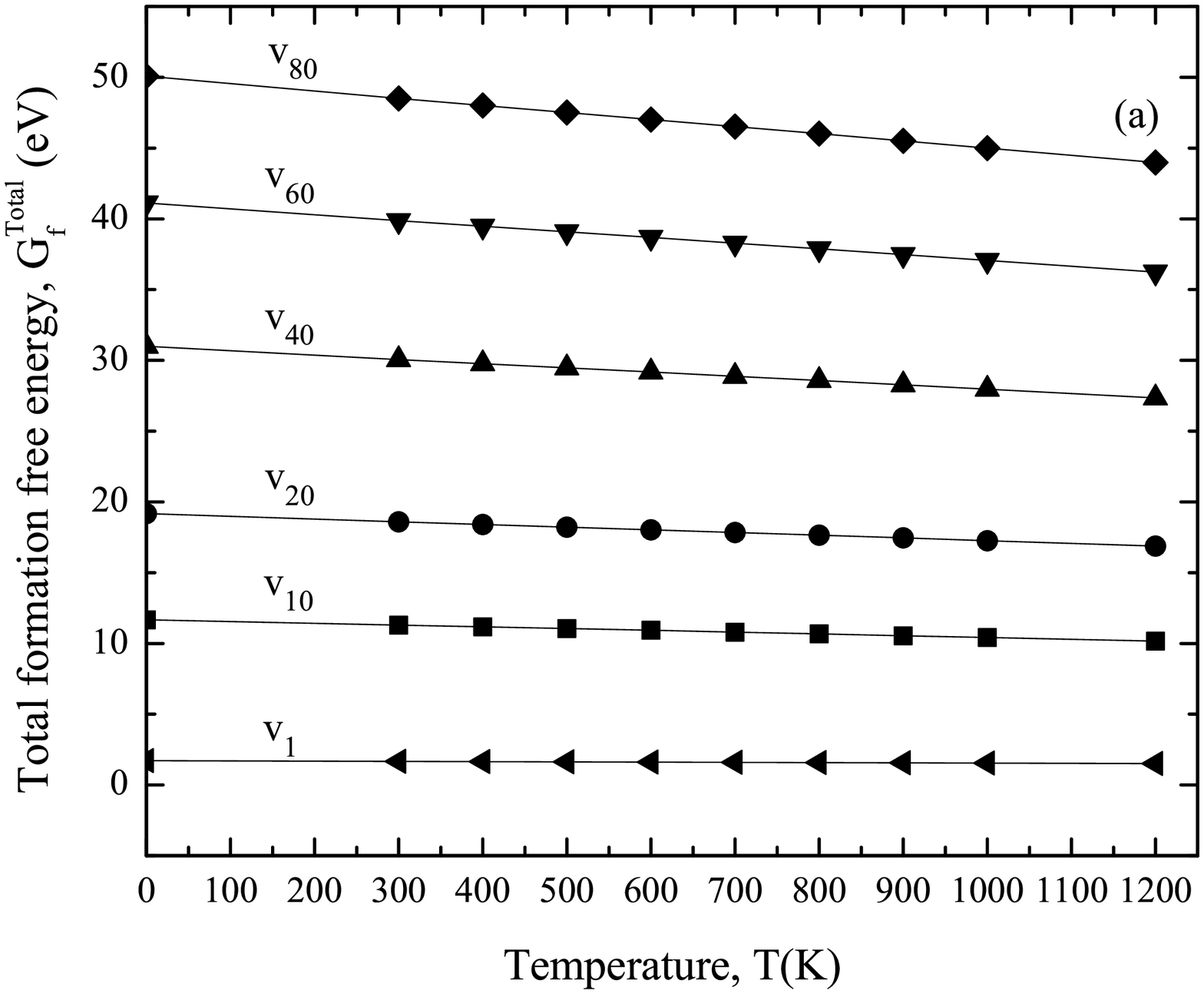}\quad
	\includegraphics[width=8cm]{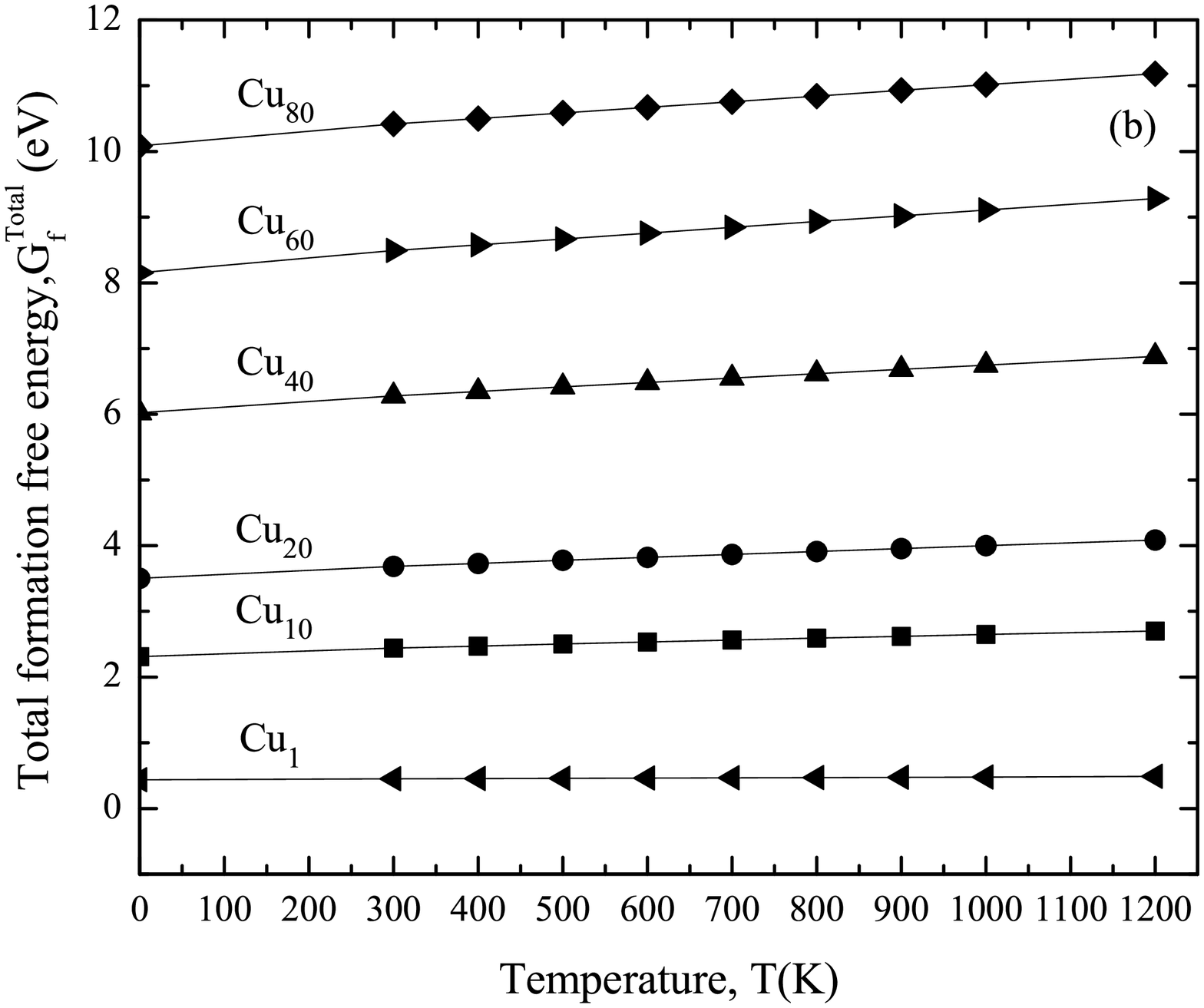}\quad
	\includegraphics[width=8cm]{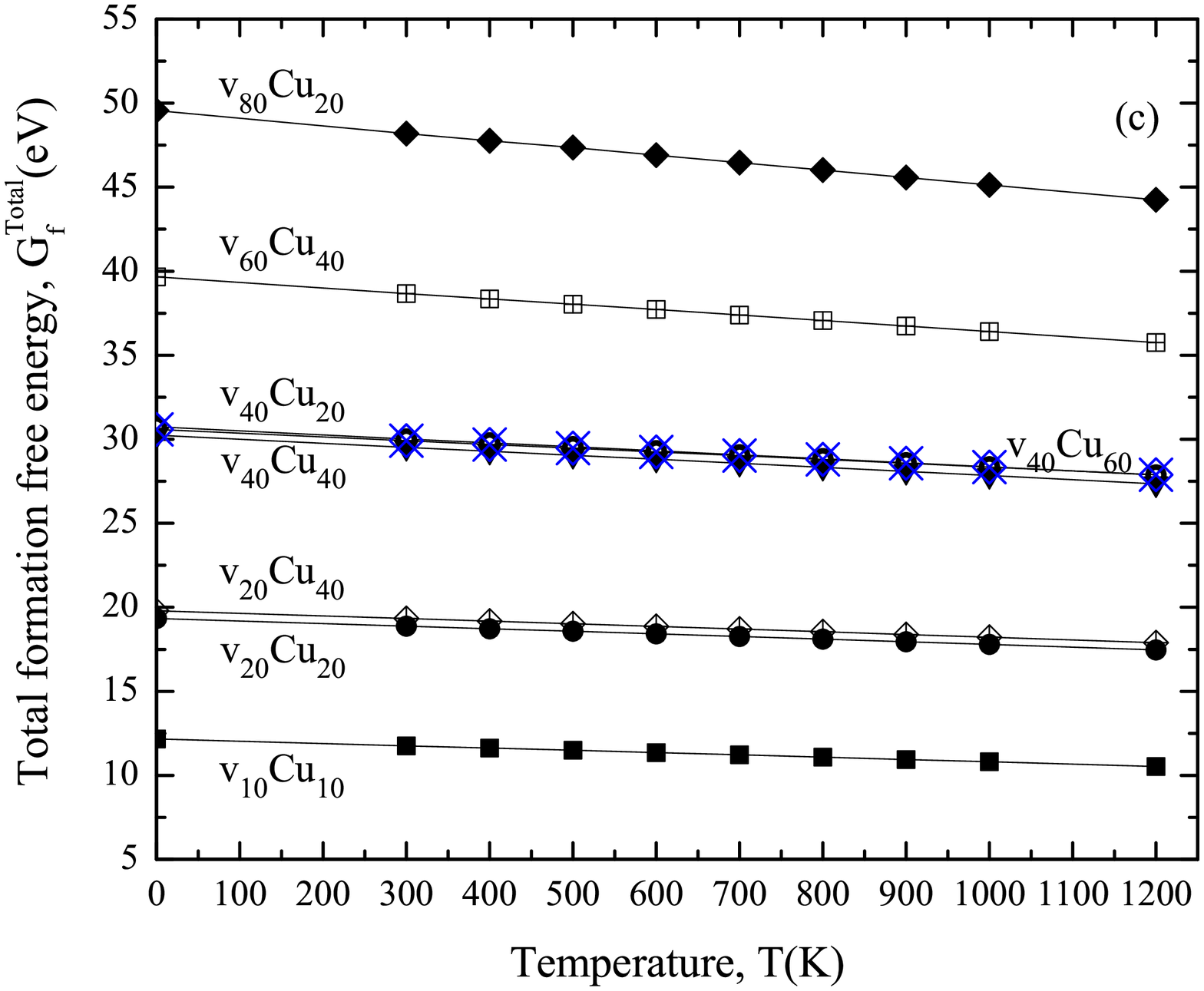}	
	\caption{\label{figs:Fig.5}Total formation free energy (eV) of (a) pure vacancy ($v_l$) clusters (b) pure Cu ($Cu_m$) clusters and (c) mixed vacancy-Cu ($v_lCu_m$) clusters in bcc-Fe.}
\end{figure}
Nevertheless, after confirming the similar trend of total formation free energy of a single Cu-atom in bcc-Fe calculated with aforemetioned two different approaches namely, the harmonic approximation and the full thermodynamic integration, we further extended our studies for pure Cu-clusters of larger sizes. The total formation free energy of pure Cu-clusters ($Cu_m$) increases with temperature within a range of 2 eV over the considered temperature range as shown in Fig. 5(b). This increase in $G^{Total}_f(Cu_m,V,T)$ means that formation or substitution of Cu-clusters in bcc-Fe matrix becomes somewhat more difficult at elevated temperature. 

Mixed vacancy-Cu clusters ($v_lCu_m$) as shown in Fig. 5(c) exhibit the effect of vibrational contribution (i.e. $G^{vib}_f(v_lCu_m ,V,T)$) by reducing the values of their total formation free energy with temperature. The similar effect is also observed in the case of pure vacancy clusters ($v_l$). Moreover, the number of vacancies present in $v_lCu_m$ clusters seems to affect the formation energy of $v_lCu_m$ clusters (i.e. $E_f(v_lCu_m)$) prominently, and further influences the clustering of their formation energy, $E_f(v_lCu_m)$ (cf. Fig. 5 in~\cite{almotasem2011}) and total formation free energy, $G^{Total}_f(v_lCu_m,V,T)$. The vibrational contribution to the total formation free energy of $v_lCu_m$ clusters, $G^{vib}_f(v_lCu_m,V,T)$ appears to affect the trend of $G^{Total}_f(v_lCu_m,V,T)$ as a function of temperature. The reduced values of $G^{Total}_f(v_lCu_m,V,T)$ at higher temperatures suggest the increased tendency of formation of mixed $v_lCu_m$ clusters and such mixed clusters are easier to form compared to pure $v_l$ clusters (cf. Fig. 5(a), Fig. 5(c)). In fact, in all cases discussed above, $E_f(v_lCu_m)$ primarily determines the values of $G^{Total}_f(v_lCu_m,V,T)$. The contribution from $G^{vib}_f(v_lCu_m,V,T)$ further modifies these values for temperature effects and also determines the trend of $G^{Total}_f(v_lCu_m,V,T)$ with temperatures.

\subsection{Binding free energy of the clusters}
Figure 6(a) shows that the absolute value of total free binding energy for pure $v_l$ clusters, $G^{Total}_{bind}(v_l,V,T)$ decreases with temperature. This decrease is associated to the increased vibrational contribution. The decreased absolute values of $G^{Total}_{bind}(v_l,V,T)$ indicate that the vibrational contribution weakens the bonding of pure $v_l$ clusters at elevated temperatures. In other words, pure vacancy clusters are less stable at elevated temperatures. At an operating temperature of a nuclear reactor close to 600 K, the absolute values of total binding free energy are about 6\% smaller than their corresponding values at 0 K. 

Pure $Cu_m$ clusters exhibit an increase in absolute values of both $G^{vib}_{bind}(v_lCu_m,V,T)$ (not shown here) and $G^{Total}_{bind}(v_l Cu_m,V,T)$ as shown in Fig. 6(b) with increasing temperature which indicates that vibrational contribution strengthens the bonding of pure $Cu_m$ clusters at elevated temperatures. This increase also means that in constrast with pure vacancy clusters, pure Cu clusters are more stable at higher temperatures. The absolute values of total binding free energy at about 600 K, are observed to increase by about 3-6\% compared to $E_{bind}(v_lCu_m)$ at 0 K. In the present study the absolute value of total free binding energy of a small $Cu_{10}$ cluster, $G^{Total}_{bind}(Cu_{10},V,T)$ calculated at 800 K shows a good agreement with the first principle calculations reported at 773 K~\cite{yuge2005}. 

As shown in Fig. 6(c) the absolute values of $G^{Total}_{bind}(v_lCu_m,V,T)$ of all mixed vacancy-copper clusters considered here decrease with increasing temperature. At 1200 K the decrease ranges from $\sim$ 1 eV for $v_{10}Cu_{10}$ clusters to $\sim$ 12 eV for $v_{80}Cu_{20}$. The $v_{20}Cu_m$, $v_{40}Cu_m$ and $v_{60}Cu_m$ clusters exhibit a decrease of $\sim$ 3-5 eV, $\sim$ 6-8 eV and $\sim$ 8-10 eV, respectively at 1200 K, where $m$ = 20, 40, 60. This reduction is primarily attributed to the increase in $G^{vib}_{bind}(v_lCu_m,V,T)$ (not shown here) and appears to follow the trend of the total binding free energy of pure $v_l$ clusters with increase in temperature. In other words, the decrease in the absolute values of total binding free energy as a function of temperature indicates the higher probablity of dissociation of these mixed clusters into their constituents i.e. \textit{l} monovacancies and \textit{m} single Cu-atoms. At about 600 K, the absolute values of total binding free energy of $v_lCu_m$ clusters is observed to reduce by minimum of $\sim$ 2.5\% for $v_{10}Cu_{10}$ to maximum of about 6\% for $v_{80}Cu_{20}$ in comparison with their corresponding values at 0 K. The $v_{20}Cu_m$, $v_{40}Cu_m$ and $v_{60}Cu_m$ clusters show a reduction of $\sim$ 5-3\%, $\sim$ 6-4\% and $\sim$ 6-5\%, respectively at 600 K in the absolute values of their $G^{Total}_{bind}(v_lCu_m,V,T)$ compared to their corresponding values at 0 K. 

\begin{figure}[ht]
  \centering
	\includegraphics[width=8cm]{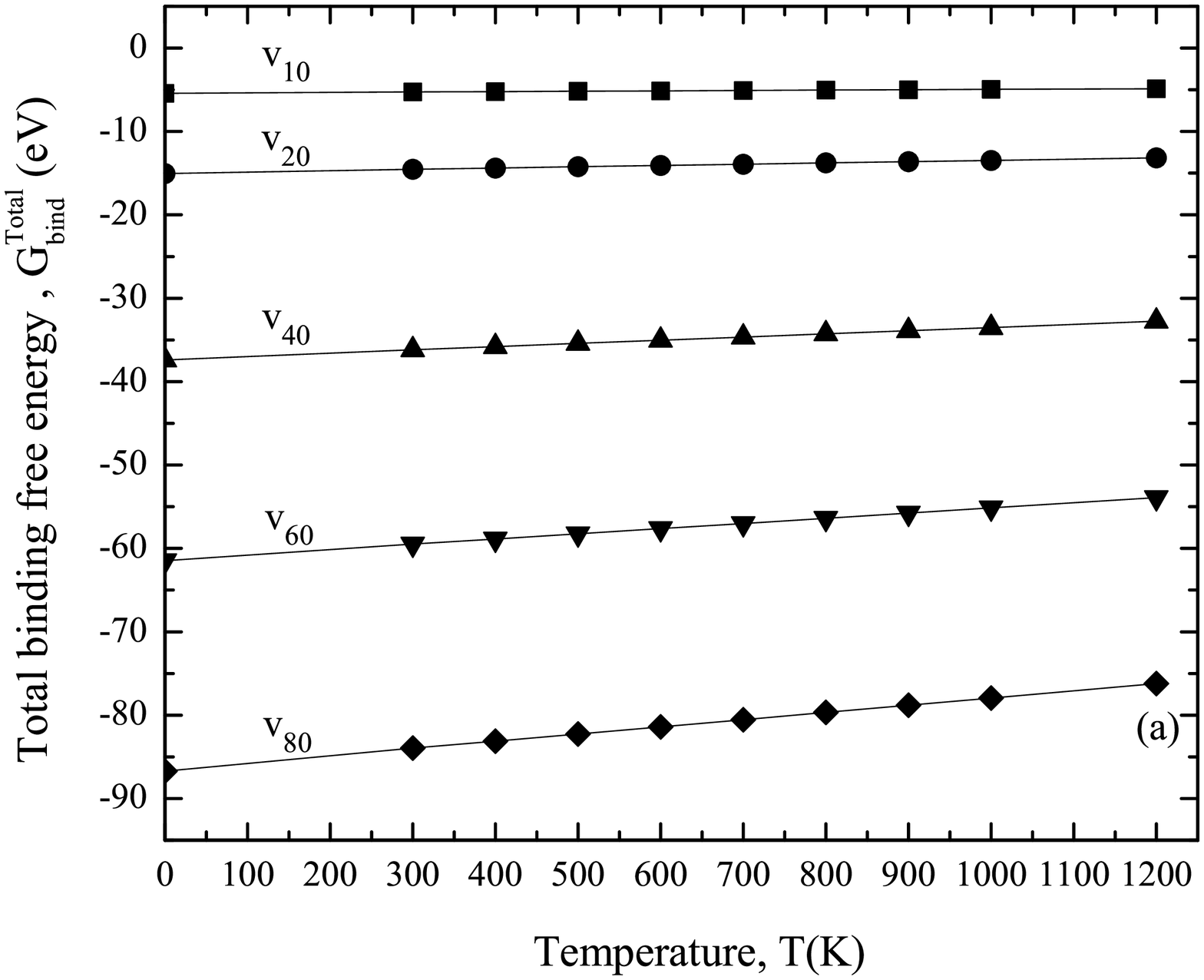}\quad
	\includegraphics[width=8cm]{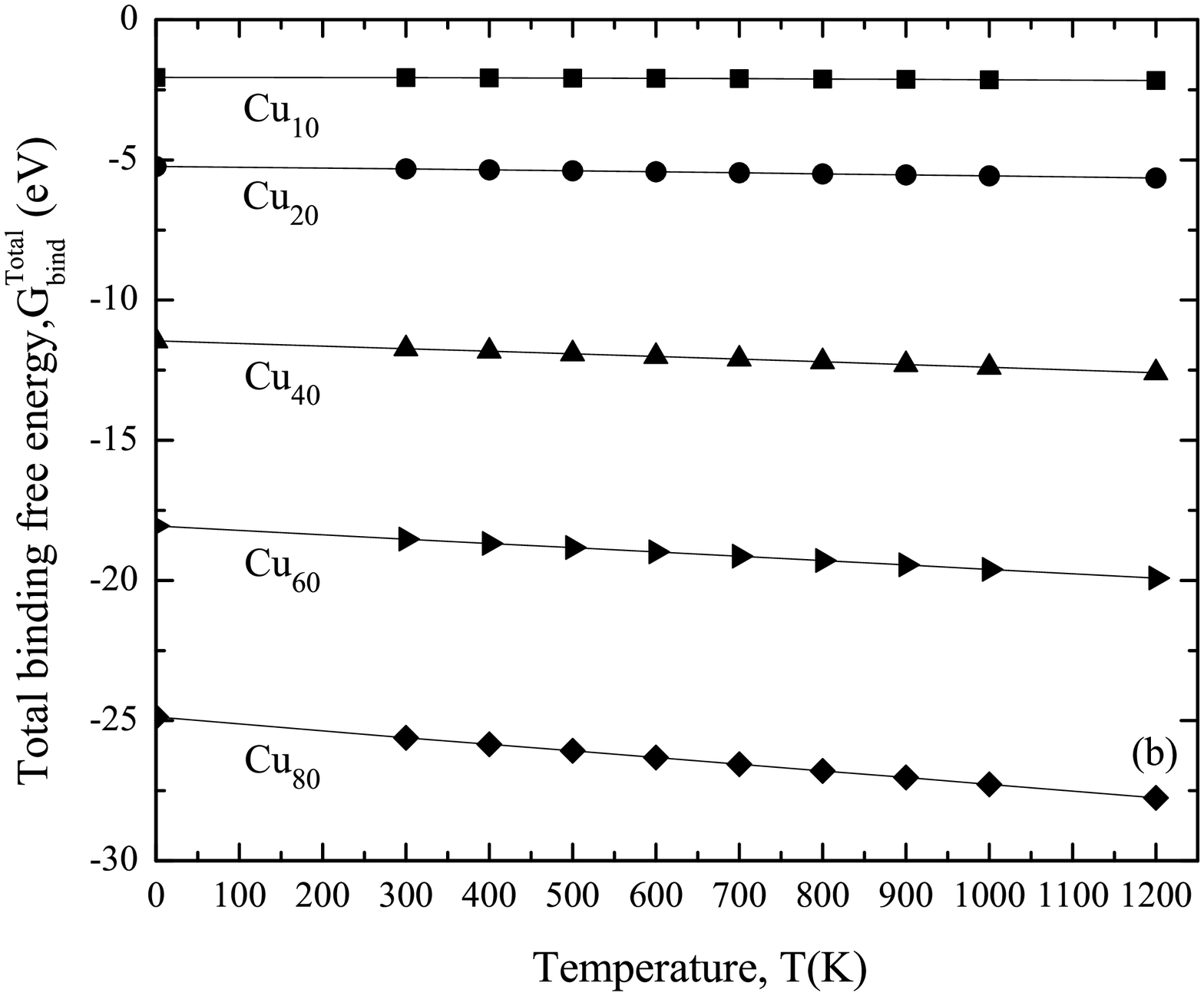}\quad
	\includegraphics[width=8cm]{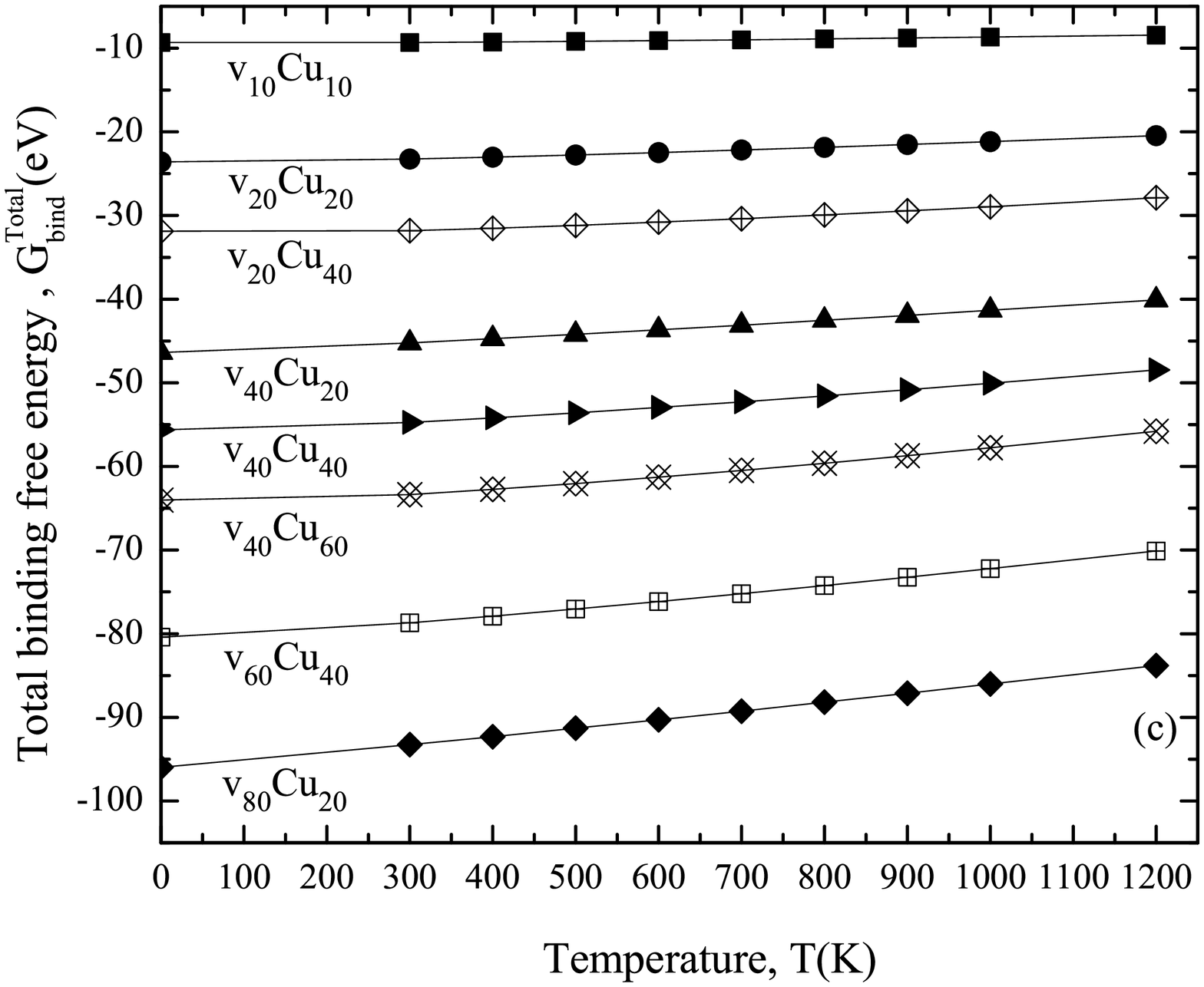}
	\caption{\label{figs:Fig.6}Total binding free energy (eV) of (a) pure vacancy ($v_l$) clusters (b) pure Cu ($Cu_m$) clusters and (c) mixed vacancy-Cu ($v_lCu_m$) clusters in bcc-Fe.}
\end{figure}

In order to estimate the monomer binding energy for arbitrary pure cluster sizes, Al-Motasem et al.~\cite{almotasem2011} have fitted the dependence of total binding energy at 0 K of pure clusters on cluster size to the following analytical relations:
\begin{eqnarray}
&& E_{bind}(v_l) \approx a\:l^{2/3} + b\:l + c\nonumber\\
&& E_{bind}(Cu_m) \approx d\:m^{2/3} + e\:m + f
\label{eq:15}
\end{eqnarray}
These fits, based on the classical capillary model, are valid for $l,m \geq 2$. The values of the fit parameters $a$ (= 2.80595 eV), $c$ (= -1.53677 eV), $d$ (= 0.59667 eV) and $f$ (= -0.60187 eV) are determined by setting $b$ and $e$ to the negative formation free energies at 0 K of the corresponding monomers, i.e. $b$ = - $E_f(v_1)$ = -1.71 eV and $e$ = - $E_f(Cu_1)$ = -0.4369 eV. The main idea to determine these analytical fits is to provide sufficiently correct and compact description of the total and the monomer binding energy for further use in rate theory and object kinetic Monte Carlo simulations. The analytical relations in Eqs.~(\ref{eq:15}) are chosen in such a way that the analytical form of their derivatives, i.e.
\begin{eqnarray}
&& E_b(v,v_l) \approx \frac{dE_{bind}(v_l)}{dl} =\frac{2}{3}a\:l^{-1/3} + b\nonumber\\
&& E_b(Cu,Cu_m) \approx \frac{dE_{bind}(Cu_m)}{dm} =\frac{2}{3}d\:m^{-1/3} + e
\label{eq:16}
\end{eqnarray}
corresponds to the relation for the monomer binding free energy used in conventinal rate theory (cf. Eq.(4) in ref.~\cite{christien2004}, for 0K). The coefficients $a$ and $d$ are related to the quantity $\gamma V^{2/3}_{at}(36\pi)^{1/3}$ given in ref.~\cite{christien2004}, where $\gamma$ denotes the cluster-matrix interface energy and $V_{at}$ the atomic volume in the bcc-Fe.

Following the concept of Al-Motasem et al., we include here the temperature effects taking vibrational part of total binding free energy of pure clusters into consideration. The analytical expressions given by Eq.~(\ref{eq:15}) are then modified for temperature effects. They are re-written as 
\begin{eqnarray}
&& G^{Total}_b(v_l,V,T) \approx a'(T)\:l^{2/3} + b'(T)\:l + c'(T)\nonumber\\
&& G^{Total}_b(Cu_m,V,T) \approx d'(T)\:m^{2/3} + e'(T)\:m + f'(T)
\label{eq:17}
\end{eqnarray}
The modified fit parameters are as follows: 
\begin{eqnarray}
&& a'(T) = 2.76602 - 2.69198 \times 10^{-4}\:T\nonumber\\
&& b'(T) = -G^{Total}_f(v_1,V,T)= -1.70968 + 1.70404 \times 10^{-4}\:T\nonumber\\
&& c'(T) = -1.18921 + 1.07434 \times 10^{-4}\:T - 3.56495 \times 10^{-8}\:T^2\nonumber\\
&& d'(T) = 0.56749 - 4.38711 \times 10^{-5}\:T\nonumber\\
&& e'(T) = -G^{Total}_f(Cu_1,V,T)= -0.4382 - 4.0428 \times 10^{-4}\:T\nonumber\\
&& f'(T) = -0.47811 + 1.82043 \times 10^{-4}\:T - 4.02524 \times 10^{-8}\:T^2\nonumber.
\end{eqnarray}

The monomer binding free energy at T $\neq$ 0 and zero pressure defined in Eq.~(\ref{eq:13}) can then be determined by treating Eq.~(\ref{eq:17}) with the similar analytical approximations introduced in Eq.~(\ref{eq:16}).  
\section{Summary and Conclusion}
In this paper, we report our work on vibrational contributions to the thermodynamics of nano-sized vacancy-copper clusters in bcc-iron. The most recent many-body empirical potential developed by Pasianot and Malerba is employed. Vibrational effects are calculated under the harmonic approximation. We summarize our results as follows:

\begin{enumerate}
	\item The vibrational free energy of bcc-Fe and fcc-Cu compares well with SGTE database and data from the other empirical potentials.	
	\item The total formation free energy of a monovacancy in bcc-Fe shows good agreement with the work of Mendelev et al. \cite{mendelev2009}. The reducing values of the total formation free energy for a monovacancy and pure vacancy clusters associated with increased vibrational effects at high temperatures exhibit an enhanced activity of vancancy formation in bcc-Fe at elevated temperatures. 
	\item The total formation free energy of a single Cu atom in bcc-Fe contradicts the predictions of the first principle calculations and it is not clear to us whether the observed results is real or an artefact of the PM potential. The increase in the total formation free energy with temperature associated with increased vibrational contribution for a single Cu atom and pure Cu clusters indicates the increased difficulty of substitution of Cu atoms in bcc-Fe. 
	\item The total formation free energy of mixed vacancy-copper clusters decreases with temperature due to increased vibrational contributions following a similar trend of pure vacancy clusters. For the number of vacancies and copper atoms considered in mixed clusters in the present work, vacancies appear to influence significantly the grouping of the total formation free energy of the mixed clusters. 
	\item The observed decrease in the absolute values of total binding free energy of pure vacancy clusters with increasing temperature indicates a higher probability of the dissociation of pure vacancy clusters into individual vancancies. The absolute values of the total binding free energy of pure vacancy clusters show the decrease of about 6\% at an operating temperature of a nuclear reactor of about 600 K as compared to their corresponding values at 0 K.
	\item The observed increase in the absolute values of the total binding free energy of pure Cu clusters on the other hand
suggests a lower probability of the dissociation of the pure Cu clusters into individual components at increased temperatures. The absolute values of the total binding free energy of pure Cu clusters show the increase of about 3-6\% at about 600 K as compared to their corresponding values at 0 K.  
	\item The absolute values of the total binding free energy of mixed defect clusters follow the trend of pure vacancy clusters and show a decrease with increasing temperature. In case of mixed clusters with considered number of vacancies and copper atoms here, vacancies overbalance vibrational contributions from copper atoms and exhibit an overall decrease in the absolute values of total binding free energy suggesting the higher probability of dissociation of mixed defects clusters into their individual constituents at higher temperatures. The absolute values of the total binding free energy of mixed clusters show a decrease from a minimum of $\sim$ 2.5\% to a maximum of $\sim$ 6\% depending on the number of vacancies present in the clusters.  
  \item Finally, the temperature dependence is included in the analytical fits in order to provide sufficiently correct and compact description of the monomer binding energy for further use in the rate theory and the object kinetic Monte Carlo simulations.
\end{enumerate}

\begin{acknowledgments}
Authors acknowledge the partial support by the European Project PERFORM60.
\end{acknowledgments}

\end{document}